\documentclass[12pt,twoside,pointlessnumbers,smallheadings]{article}

\input psfig.sty

\newcommand{\psip}{\psi(2S)}

\newcommand{\jpsi}{J/\psi}

\newcommand{\EE}{e^+e^-}
\newcommand{\MM}{\mu^+\mu^-}
\newcommand{\TT}{\tau^+\tau^-}
\newcommand{\GG}{\gamma\gamma}

\newcommand{\beq}{\begin{equation}}
\newcommand{\eeq}{\end{equation}}
\newcommand{\beqn}{\begin{eqnarray}}
\newcommand{\eeqn}{\end{eqnarray}}
\newcommand{\beqns}{\begin{eqnarray*}}
\newcommand{\eeqns}{\end{eqnarray*}}
\newcommand{\bfg}{\begin{figure}}
\newcommand{\efg}{\end{figure}}
\newcommand{\bitm}{\begin{itemize}}
\newcommand{\eitm}{\end{itemize}}
\newcommand{\bnum}{\begin{enumerate}}
\newcommand{\enum}{\end{enumerate}}
\newcommand{\btbl}{\begin{table}}
\newcommand{\etbl}{\end{table}}
\newcommand{\btbu}{\begin{tabular}}
\newcommand{\etbu}{\end{tabular}}
\begin{document}          

\small

\title{Determination of $\psip$ Total Number by Inclusive Hadronic
Decay\thanks{
Supported by National Natural Science Foundation of China (19991483)
and 100 Talents Program of CAS (U-25)} }

\author{MO Xiao-Hu$^{1, 2; }$~\footnote{E-mail:moxh@mail.ihep.ac.cn}
 ~~ZHU Yong-Sheng $^{1}$ ~~YUAN Chang-Zheng $^{1}$ \\
 ~~FU Cheng-Dong $^{1}$  ~~Li Gang $^{1}$\\
{\small 1 ( Institute of High Energy Physics, CAS,
Beijing 100039, China ) ~~~~~~~~~~~~~~~} \\
{\small 2 ( China Center of Advanced Science and Technology,
Beijing 100080, China )} }

\date{October 10, 2003}
\maketitle

\begin{center}
\begin{minipage}{15cm}
{\small
{\bf Abstract} \hskip 0.25cm
On the basis of the study of inclusive hadronic events, two methods are adopted 
to determine the number of produced $\psip$ events collected by BES in 
2001$-$2002 run, which is $14.0 \times 10^6$ with the uncertainty of 4\%.

 }
\end{minipage}
\end{center}

\section{Introduction}

The BEijing Spectrometer (BES) is a general purpose solenoidal 
detector$^{\mbox{\small \cite{detector}}}$ running at Beijing Electron Positron Collider (BEPC). 
The beam energy of BEPC is in the range from 1.5 GeV to 2.8 GeV with a 
design luminosity of $1.7 \times 10^{31} \mbox{cm}^{-2} \mbox{s}^{-1}$ at 5.6 
GeV center of mass energy. The main physics goal is to study the charm and 
$\tau$ physics. During 2001-2002 years' running, about 14 million $\psip$ online hadronic events have been collected\footnote{Totally 3000 runs 
(RUN20050-RUN23085) are taken, among which
only those with $I_{quality}=2,3$ are used to determine the total number of
the $\psip$ events. Here $I_{quality}$ denotes the run quality and value 2 and 3 indicate that the run quality is fairly reliable.}.
On the basis of this large data sample, many physics analyses could be
performed with an unprecedented precision.

The determination of the offline total number of $\psip$ event,
$N^{TOT}_{\psip}$, is a foundational work in physics analysis, and in turn is 
the foundation of the further analysis study. In $\psip$ physics analysis, the calculation of the absolute branching ratio depends on $N^{TOT}_{\psip}$, whose error will be directly accounted into the error of the branching ratio of
any being studied channel. Therefore, it is essential to work out the
$N^{TOT}_{\psip}$ accurately and reliably. In principle, any decay channel with known branching ratio could be used to evaluate the total number of $\psip$:
$$
N^{TOT}_{\psip} = \frac{N^{obs}_{f}}{ \epsilon_f \cdot {\cal B}_f }~,
$$
where $N^{obs}_{f}$ is observed number of final state $f$ for a certain decay 
channel, $\epsilon_f$ and ${\cal B}_f$ are the corresponding efficiency and the branching ratio. It is obvious that the larger the branching ratio and the 
smaller the corresponding error, the more reliable the total number is. On such an extent, the inclusive hadron final state is a favorable process for the 
total number determination. The only disadvantage here lies 
in the difficulty to eliminate all kinds of backgrounds throughly. Therefore 
the meticulous studies have been made for the hadron event selection.

In the following sections, the hadron event selection is discussed firstly,
then two methods are utilized to determine the total number of $\psip$ and
the uncertainties from various sources are studied. At last, the final result 
is given.

For clearness and convenience, some notations which are to be used afterwards, 
are listed in the Table \ref{notation}.

\begin{table}[htb]
\caption{\label{notation} Notations}
\vskip 0.2cm
\center
\begin{tabular}{c|c||c|c||c|c} \hline \hline 
   Symbol &     Meaning   &Superscript& Meaning &Subscript& Meaning \\ \hline
$    N   $&Observed Number&$   T   $&Total      &$   h  $&hadron final state \\
$    m   $&Selected Number&$   P   $&Peak region&$   e  $&$\EE$ final state  \\
$    n   $&``Pure'' Hadron Number
                          &$   R   $&Resonance  &        &                   \\
$\epsilon$&Efficiency     &$   C   $&Continuum  &        &                   \\
$\sigma  $&Cross section  &                     &        &                   \\
$    L   $&Integrated Luminosity     
                          &         &           &        &\\ \hline \hline                 
\end{tabular}
\end{table}
In addition, there are two elementary relations among the five quantities 
$N$, $n$, $\epsilon$, $\sigma$, and $L$, that is
\begin{eqnarray}
{\displaystyle L=\frac{N}{\sigma} }  & \mbox{   or  } &
N=L \cdot \sigma \label{rela_1}~~,\\
{\displaystyle N=\frac{n}{\epsilon}} & \mbox{   or  } &
n=N \cdot \epsilon \label{rela_2}~~.
\end{eqnarray}
The symbol with a tilde on it ($e.g.$ $\tilde{n}$, $\tilde{N}$, $etc.$),
denotes the events obtained at the continuum region ($E_{beam}=3.665$ GeV), 
while others denote the events obtained at the resonance region ($E_{beam}=
3.686$ GeV). 

There are also two frequently used equalities, the first one for variables
of the same process at different energy points:
\begin{equation}
\frac{\tilde{N}/\tilde{\sigma}}{N/\sigma} = \frac{\tilde{L}}{L} =
\frac{\tilde{n}/(\tilde{\epsilon}\cdot \tilde{\sigma})}
     {n/(\epsilon \cdot \sigma)}~~, 
\mbox{\hskip 0.2cm   or  \hskip 0.2cm} 
\frac{\tilde{n}}{n}=\frac{\tilde{L}\cdot \tilde{\sigma}}{L\cdot \sigma}
\mbox{ ( for $\tilde{\epsilon}=\epsilon$ )}~~;
\label{rela_A}
\end{equation}
the second one for variables of different process at the same energy points:
\begin{equation}
\frac{n^I/\epsilon^I}{n^J/\epsilon^J}=
\frac{N^I}{N^J}=\frac{L\cdot \sigma^I}{L\cdot \sigma^J}
=\frac{\sigma^I}{\sigma^J}~~,
\mbox{\hskip 0.2cm   or  \hskip 0.2cm}
\frac{n^I}{n^J}=
\frac{\epsilon^I \cdot \sigma^I}{\epsilon^J \cdot \sigma^J}~~.
\label{rela_B}
\end{equation}

\section{Hadron Event Selection}\label{set_hadsel}

For the hadron event selection, the detail information could be found in 
Refs.~\cite{psipscan} and~\cite{hadst}. There is no particular event 
topology to require; instead cuts are made to reject major backgrounds:
cosmic rays, beam associated backgrounds, two-photon process ($\gamma^{\ast}
\gamma^{\ast}$), mis-identified ``hadron'' event from QED processes of
$\EE \rightarrow l^+l^-, l=e,\mu,\tau$, and $\EE \rightarrow \GG$
followed by $\gamma$ conversion, and so forth. Most of these kinds of event
have salient topology and could be eliminated by proper criteria.
Events with at least two well reconstructed charged tracks within
$|\cos \theta| \leq 0.8$ are selected (that is $N_{good} \geq 2$). The total
energy deposited by an event in the BSC ($E_{sum}$) is required to be larger
than 0.36 $E_{beam}$, in order to suppress the contamination from two-photon 
processes and beam associated backgrounds. Events with all tracks pointing to
the same hemisphere in at least one of axial directions ($x$ or $y$ or $z$
direction) are removed to suppress beam associated backgrounds. (This 
requirement could be expressed quantitatively as $I_{ssi} \geq 1 $, where
$I_{ssi}$ is called the squared spatial distribution index.) For two-prong
events, two additional cuts are applied to eliminate possible lepton pair
backgrounds. The number of photons must be greater than one (that is 
$N_{real~\gamma } \geq 2$), and the acollinearity between two charged tracks,
$\alpha_{Acol}$, must be greater than 10 degrees. 

\begin{figure}[hbt]
\begin{minipage}{7cm}
\centerline{\psfig{file=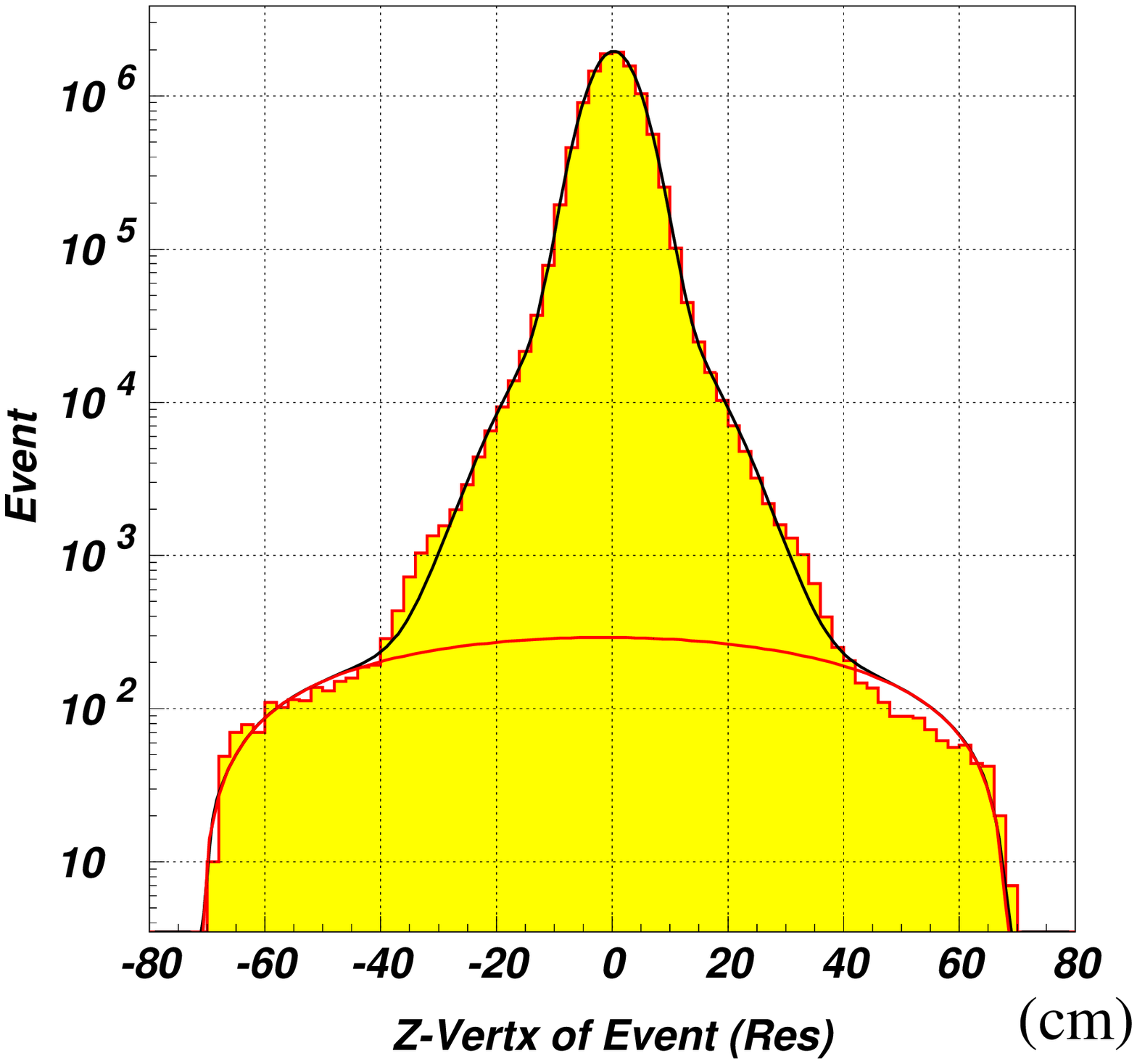 ,height=6.5 cm ,width=7.5 cm}}
\caption{\label{fsh2r} Vertex-Fit distribution ($E_{cm}=3.686$ GeV).}
\end{minipage}
\hskip 0.5cm
\begin{minipage}{7cm}
\centerline{\psfig{file=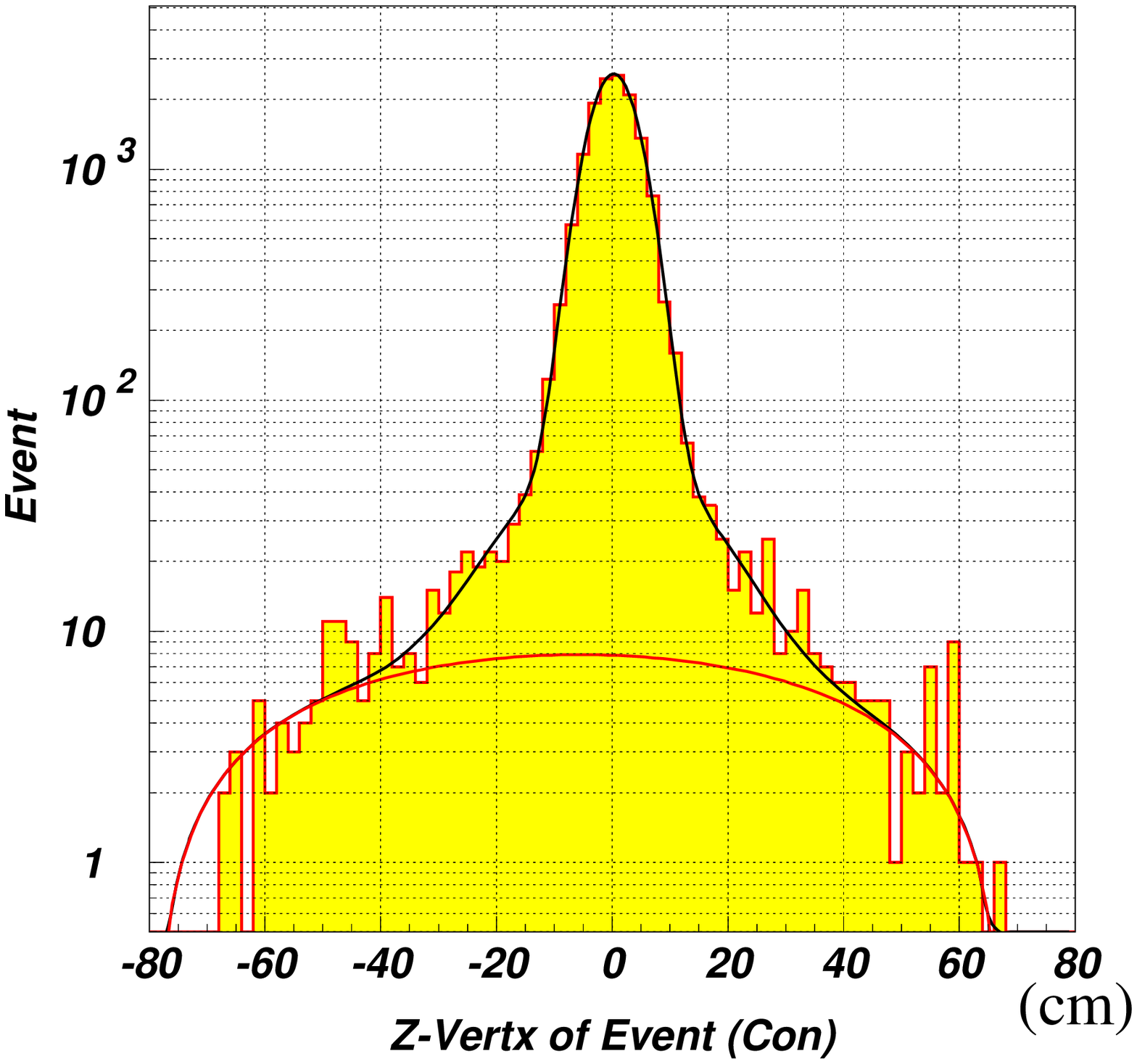 ,height=6.5 cm ,width=7.5 cm}}
\caption{\label{fshc} Vertex-Fit distribution ($E_{cm}=3.665$ GeV).}
\end{minipage}
\end{figure}

After the event selection, the fitting of double Gaussian plus a polynomial is 
applied to eliminate the remained background from beam associated backgrounds,
see Fig.~\ref{fsh2r} and \ref{fshc}.

The global error analysis approach is adopted to obtain the uncertainties of
selection cuts~\cite{hadst}, as listed in Table~\ref{haderr}:

\begin{table}[bth]
\doublerulesep 0.5pt
\caption{\label{haderr} Error of hadron event selection.}
\center
\begin{tabular}{cc} \hline \hline
       Requirement        &    Error       \\ \hline \hline
  $N_{good} \geq 2 $      &    2.765 \%    \\
 $E_{sum}>0.36E_{beam}$   &    2.544\%     \\
       $I_{ssi}< 1 $      &    0.173 \%    \\
  $N_{real\gamma}\geq 2$
(for $N_{good}= 2$ )      &    0.042 \%    \\
  $\alpha_{Acol}\geq 10^{\circ} $
  (for $N_{good}= 2$ )    &    0.157 \%    \\  \hline \hline
         Sum              &    3.765 \%    \\  \hline \hline
\end{tabular}
\end{table}

\begin{figure}[hbt]
\centerline{\psfig{file=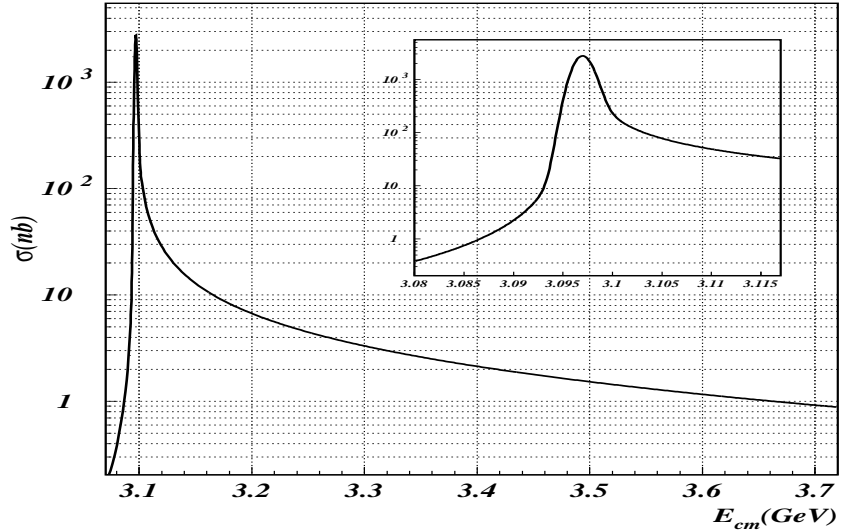,width=11.0cm,height=7.0 cm}}
\caption{\label{jpsect} Cross section curve of $J/\psi$ resonance.}
\end{figure}

In fact, there are many processes that could lead to hadron final state 
at $\psip$ resonance region, they could be divided into seven categories:
\begin{eqnarray}
\EE &\rightarrow &hadron  \mbox{~~~(CH)}~~, \label{conhad} \\
\EE &\rightarrow &\psip \rightarrow hadron
                          \mbox{~~~(RH)}~~, \label{reshad} \\
\EE &\rightarrow &\jpsi \rightarrow hadron
                          \mbox{~~~($\jpsi$-H)}~~, \label{jpsihad} \\
\EE &\rightarrow &\TT \rightarrow hadron 
                          \mbox{~~~($\tau$-CH)}~~, \label{tauchad} \\
\EE &\rightarrow &\psip \rightarrow \TT \rightarrow hadron 
                          \mbox{~~~($\tau$-RH)}~~, \label{taurhad} \\
\EE &\rightarrow & hadron^{\ast} 
                          \mbox{~~~(CH$^\ast$)}~~, \label{xchads} \\
\EE &\rightarrow &\psip \rightarrow hadron^{\ast} 
                          \mbox{~~~(RH$^\ast$)}~~, \label{xrhads} 
\end{eqnarray}
where C represents the continuum process, R the resonance process, 
H hadron event, and H$^\ast$ indicates the event which survives all 
aforementioned hadron selection cuts and is left in hadron sample from processes (\ref{xchads}) and (\ref{xrhads}). 
Among above seven categories, only the hadron event from the first 
two processes are ``pure'' 
hadron event at $\psip$ peak region while others should be treated as
backgrounds. Since hadron event from different process has almost the same event
topology, the theoretical estimation method is used to evaluate the 
contamination of such kinds of backgrounds. According to the analysis in 
Ref.~\cite{hadst}, two factors are introduced to subtract the hadronic 
background. If the pure hadron number is denoted as $n$ and the selected hadron number denoted as $m$, then it could be obtained 
\begin{eqnarray}
 \gamma_C m^C_{had} &=& n^C_{had}~~,\label{mncon}  \\
 \gamma_R m^R_{had} &=& n^R_{had}~~,\label{mnres}
\end{eqnarray}
where
\begin{eqnarray*}
\gamma_C &=& 1 - f_{\jpsi} - f_C = 0.855~~, \label{faccon} \\
\gamma_R &=& 1 - f_R  =0.999~~,\label{facres}
\end{eqnarray*}
with
$$
f_{\jpsi}= \frac{\sigma^{\jpsi}_{had} \cdot \epsilon^{\jpsi}_{had}}
                {\sigma^{C}_{had} \cdot \epsilon^{C}_{had}}~~,
$$
$$
f_C = \sum\limits_{k=\tau,e,\mu,\gamma,\gamma^*} 
        \frac{\sigma^{C}_k \cdot \epsilon^C_k}
             {\sigma^{C}_{had} \cdot \epsilon^{C}_{had}}~~,
$$
$$
f_R = \sum\limits_{l=\tau,e,\mu,\gamma} \frac{B^{\psip}_l}{B^{\psip}_h} \cdot
\frac{\epsilon^R_l}{\epsilon^R_h}~~.
$$
Here all efficiencies are obtained from Monte Carlo simulation~\cite{hadst},
and symbols $\tau,e,\mu,\gamma$ and $\gamma^*$ denote final states 
$\TT,\EE,\MM,\GG$ and $\gamma^* \gamma^*$, respectively. For the continuum 
process, the production cross section $\sigma^C_k$ could be obtained from the 
corresponding Monte Carlo generator. For $\sigma^C_{had}$ and 
$\sigma^{\jpsi}_{had}$, they could be calculated by theoretical formulae 
with corresponding resonance parameters obtained from the scan experiment
data~\cite{psipscan,jpsiscan}, 
which is about 15 nb and 1 nb, respectively. Since Monte Carlo generator does 
not give the cross section for resonance process, the ratio of branching 
fractions is used in the calculation of factor $f_R$~\cite{hadst}. It should be
pointed out that because the variation of $\sigma^{\jpsi}_{had}$ is fairly
smooth at $\psip$ region, refer to Fig.~\ref{jpsect}, the contamination 
from $\jpsi$ decay is suitable to be treated as the continuum-like hadron
background.

\section{Determination of Total Number}
\subsection{Principle}

\begin{figure}[hbtp]
\centerline{\psfig{file=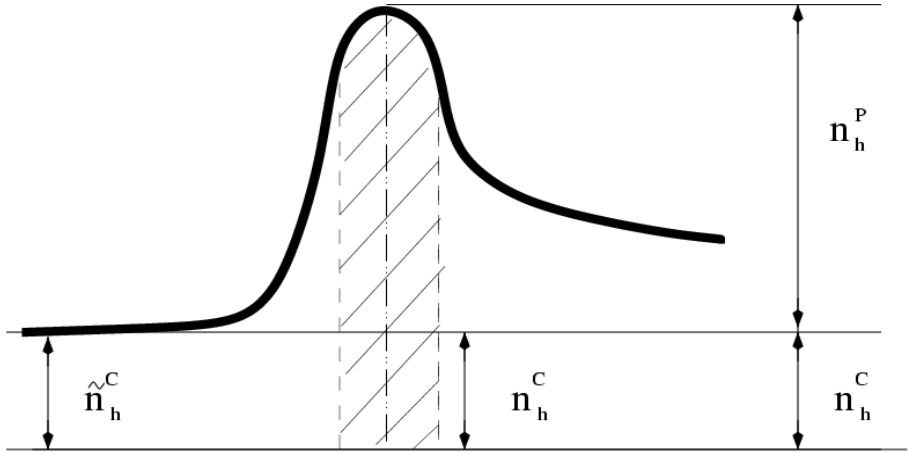 ,height=6.0cm ,width=12.0cm}}
\caption{\label{hadraw}Fractional sketch of hadron number of
different process. (The cross-hatched area in the figure indicates the data
taking region.)}
\end{figure}

The branching ratio of hadron final state, denoted as ${\cal B}_h$,
can be acquired from PDG list~\cite{PDG} 
or from BES scan results \cite{psipscan}. If the number of selected hadron
event from $\psip$ resonance is $n_{\psip \rightarrow had.}$, then 
\begin{equation}
N^{TOT}_{\psip}=\frac{ n_{\psip \rightarrow had.} }
                     { \epsilon^R_h \cdot {\cal B}_h }~~.
\label{totnmb}
\end{equation}
However, the number of selected hadron event at a certain energy point is the
combination of two parts (refer to Fig.~\ref{hadraw}), one from resonance 
process and the other from continuum one, that is\footnote{\label{ftnt_par}
Because of energy spread and shift, the data are actually taken within a
region rather than at a fixed point (notice the cross-hatched area in 
Fig.~\ref{hadraw}). So ``$P$'' instead of ``$R$'' is adopted
in the following formulae in order to denote such an effect.}
\begin{equation}
\begin{array}{ccccc}
n_{had}   &= & n_{\psip \rightarrow had.} &+ & n_{\EE \rightarrow had.}~~, \\
\Downarrow&  & \Downarrow                 &  & \Downarrow \\
n^T_{h}   &= & n^R_h (n^P_h)              &+ & n^C_h~~ .
\end{array}
\label{eq_trc}
\end{equation}
Therefore the key issue here is to distinguish the $n^R_h$ from the $n^T_h$. 
There are two methods, the fraction subtraction method and the normalization 
subtraction method, can be used to figure out the number of resonance event.

\subsection{Fraction subtraction} 
Refer to Fig. \ref{hadraw}, if the ratio of $n^C_h$ to $n^P_h$ could
be estimated:
$$
f=\frac{n^C_h}{n^P_h}~~,
$$
together with Eq.~(\ref{eq_trc}), the $n^P_h$ can be calculated as
\begin{equation}
n^P_h = \frac{n^T_h}{1+f}~~.
\label{nmbp1}
\end{equation}
As to the factor $f$, from Eq.~(\ref{rela_B}), it is easy to acquire
$$
\frac{n^C_h}{n^P_h} =
\frac{\epsilon^C_h \cdot \sigma^C_h}{\epsilon^R_h \cdot \sigma^R_h}~~,
\mbox{\hskip 0.2cm   or  \hskip 0.2cm}
n^C_h = \frac{\epsilon^C_h \cdot \sigma^C_h}{\epsilon^R_h \cdot \sigma^R_h}
\cdot n^P_h~~,
$$
so $f$ can be expressed as
\begin{equation}
f=\frac{\epsilon^C_h \cdot \sigma^C_h}{\epsilon^R_h \cdot \sigma^R_h}~~.
\label{facf1}
\end{equation}
Combing Eqs. (\ref{totnmb}) and (\ref{nmbp1}), it can be obtained 


\begin{equation}
N^{TOT}_{\psip}=\frac{n^T_h}{{\cal B}_h \cdot \epsilon^R_h \cdot (1+f)}~~.
\label{totnmb1}
\end{equation}

\subsection{Normalization subtraction}
The data taken far from the resonance region could be treated as the
data of continuum\footnote{Strictly speaking, within the scan range, any data 
are from two processes, resonance and continuum. The effect of resonance to 
continuum could be taken into account by factor $f_e, \tilde{f}_e, \tilde{f}_h$ in Eqs. (\ref{fact_e}), (\ref{fact_te}), and (\ref{fact_th}).}, 
refer to Fig. \ref{hadraw}, from Eq.~(\ref{rela_A}), the number at continuum 
region $\tilde{n}^C_h$ could be transformed into that at resonance $n^C_h$ by a luminosity normalization factor
$$
n^C_h = f_T \tilde{n}^C_h~~, \hskip 2cm
f_T = \frac{\sigma^C_h \cdot L}{\tilde{\sigma}^C_h \cdot \tilde{L}}~~,
$$
where $f_T$ is the transformed factor. Combining with the relation
$ n^T_h= n^P_h + n^C_h$, the resonance number could be worked out
$$
n^P_h = n^T_h - f_T \cdot \tilde{n}^C_h 
      = n^T_h \cdot (1 - f_T \cdot \frac{\tilde{n}^C_h}{n^T_h} )~~.
$$
In the expression of $f_T$, the luminosity $L$ is usually calculated by 
the continuum $\EE$ event ($i.e.$ Bhabha event):
$$
L=\frac{n^C_e}{\epsilon^C_e \cdot \sigma^C_e}~~~.
$$
Similar to Eq.~(\ref{nmbp1}), at the continuum region, the number of events 
from the continuum process is expressed by the number of total selected events:
$$
n^C_e=\frac{n^T_e}{1+f_e}~~, \hskip 0.5cm
\tilde{n}^C_e=\frac{\tilde{n}^T_e}{1+\tilde{f}_e}~~, \hskip 0.5cm
\tilde{n}^C_h=\frac{\tilde{n}^T_h}{1+\tilde{f}_h}~~,
$$
with
\begin{eqnarray}
f_e ={\displaystyle 
\frac{\epsilon^R_e \cdot \sigma^R_e}{\epsilon^C_e \cdot \sigma^C_e} }~~,
\label{fact_e} \\
\tilde{f}_e ={\displaystyle
             \frac{\epsilon^R_e \cdot \tilde{\sigma}^R_e}
                  {\epsilon^C_e \cdot \tilde{\sigma}^C_e} }~~,
\label{fact_te} \\
\tilde{f}_h ={\displaystyle
             \frac{\epsilon^R_h \cdot \tilde{\sigma}^R_h}
                  {\epsilon^C_h \cdot \tilde{\sigma}^C_h } }~~.
\label{fact_th}
\end{eqnarray}
It should be noticed that 
the relation $\epsilon^{R,C}_{e,h}=\tilde{\epsilon}^{R,C}_{e,h}$ 
has been used\footnote{
The relation $\epsilon^{R,C}_{e,h}=\tilde{\epsilon}^{R,C}_{e,h}$ is
exact for the $\EE$ final state, whose event selection cuts are energy
independent; but for the $hadron$ final state, the relation is only an
approximation.} in the above calculation.

Put all together, the total number can be calculated as
\begin{equation}
N^{TOT}_{\psip}=\frac{n^T_h}{{\cal B}_h \cdot \epsilon^R_h}
\cdot \left( 1- F_T \cdot M_{eh} \right) ~~, 
\label{totnmb2}
\end{equation}
where 
\begin{equation}
F_T \equiv \frac{\sigma^C_h \cdot \tilde{\sigma}^C_e}
                {\tilde{\sigma}^C_h \cdot \sigma^C_e}
\cdot \frac{(1+ \tilde{f}_e)}{(1+ f_e)} 
\cdot \frac{1}{(1+ \tilde{f}_h)}~~,
\label{ftfirst}
\end{equation}
\begin{equation}
M_{eh}= \frac{n^T_e}{\tilde{n}^T_e} \cdot \frac{\tilde{n}^T_h}{n^T_h}~~,
\end{equation}
with factors $f_e$, $\tilde{f}_e$, and $\tilde{f}_h$ are given in
Eqs. (\ref{fact_e}), (\ref{fact_te}) and (\ref{fact_th}).

\subsection{Correction}
As mentioned in section~{\bf \ref{set_hadsel}}, after the hadron event 
selection, the selected hadron number $m$ instead of pure one $n$ is obtained. The relation between $m$ and $n$ is given in Eq.~(\ref{mncon}) and
~(\ref{mnres}), that is
\begin{eqnarray*}
 \gamma_C (m^C_{h},\tilde{m}^C_{h}) &=& (n^C_{h},\tilde{n}^C_{h})~~,  \\
 \gamma_R (m^R_{h},\tilde{m}^R_{h}) &=& (n^R_{h},\tilde{n}^R_{h})~~.
\end{eqnarray*}

Notice that
$$
m^T_{h} = m^P_h + m^C_h~~,~~~~~(m^P_h = m^R_h )~~, 
$$
then
$$ m^T_{h} = \frac{n^P_{h}}{\gamma_R}+ \frac{n^C_{h}}{\gamma_C}
 = \frac{n^P_{h}}{\gamma_R}+ \frac{f n^P_{h}}{\gamma_C}
 = n^P_{h} \cdot \left( \frac{1}{\gamma_R} + \frac{f}{\gamma_C} \right)~~, $$
that is
$$ n^P_{h} = \frac{m^T_{h}}
{\displaystyle \left( \frac{1}{\gamma_R} + \frac{f}{\gamma_C} \right)} ~~.$$
So the formula for the fraction subtraction method now becomes
\begin{equation}
N^{TOT}_{\psip} = \frac{m^T_h}
{{\cal B}_h \cdot \epsilon^R_h \cdot (1+f) \cdot \delta_{\gamma_1} }~~, 
\label{totnmb1r}
\end{equation}
where $\delta_{\gamma_1}$ is defined as
\beq
\delta_{\gamma_1} 
= \frac{\displaystyle \frac{1}{\gamma_R} + \frac{f}{\gamma_C} }
       {\displaystyle 1+ f}~~.
\label{dga1}
\eeq

For the normalization subtraction method, the corresponding corrected
formula could be obtained similarly and the final result is

\begin{equation}
N^{TOT}_{\psip}=\frac{m^T_h}{\gamma_R \cdot {\cal B}_h \cdot \epsilon^R_h}
\cdot \left( 1- {\cal F}_T \cdot {\cal M}_{eh} \right) ~~, 
\label{totnmb2r}
\end{equation}
where
\begin{equation}
{\cal M}_{eh}= \frac{n^T_e}{\tilde{n}^T_e}\cdot\frac{\tilde{m}^T_h}{m^T_h}~~,
\end{equation}
and 
\beq
{\cal F}_T = F_T / \delta_{\gamma_2}~~, ~~~~~~
\delta_{\gamma_2} \equiv 
\frac{\displaystyle 1+ \frac{\gamma_C}{\gamma_R} \tilde{f}_h}
     {1+ \tilde{f}_h}~~,
\eeq
here $F_T$ is just as that defined in Eq.~(\ref{ftfirst}).

\subsection{Numerical Calculation}
By the virtue of Eqs.~(\ref{totnmb1r}) and (\ref{totnmb2r}), $N^{TOT}_{\psip}$
could be worked out. For convenience, all numbers relevant to total number 
calculation are summarized in Table~\ref{nmbeff}. 

\begin{table}[htb]
\caption{\label{nmbeff} Event number, efficiency and cross section.}
\vskip 0.15cm
\doublerulesep 0.5pt
\begin{center}
\begin{tabular}{c|c||cc||cc} \hline \hline 
\multicolumn{2}{c||}{Final state}
         & \multicolumn{2}{|c||}{$hadron$} 
             & \multicolumn{2}{|c}{$\EE$}  \\ \hline 
\multicolumn{2}{c||}{Region}
         &Resonance               &Continuum
             &Resonance               &Continuum                \\ \hline 
\multicolumn{2}{c||}{Event number}
         &$m^T_{h}=10634586.0$        &$\tilde{m}^T_{h}=13937.7$
             &$n^T_{e}=1190613$       &$\tilde{n}^T_{e}=54415$  \\
\multicolumn{2}{c||}{$m^{fit}_{h}$ ($\Delta m^{fit}_{h}$) }
          &($\Delta m^T_{h}=11688.2$)&($\Delta \tilde{m}^T_{h}=817.2$)
             &                        &                         \\ \hline
Efficiency&Resonance
     &$\epsilon^R_h=0.753$ & $\tilde{\epsilon}^R_h \approx \epsilon^R_h$
     &$\epsilon^R_e=0.761$ & $\tilde{\epsilon}^R_e =\epsilon^R_e$ \\
          &process 
     &($\Delta \epsilon^R_h=0.012$) & & & \\ \cline{2-6}
          &Continuum
     &$\epsilon^C_h=0.716$ & $\tilde{\epsilon}^C_h \approx \epsilon^C_h$
     &$\epsilon^C_e=0.708$ & $\tilde{\epsilon}^R_e =\epsilon^R_e$ \\ 
          &process 
     &($\Delta \epsilon^R_h=0.010$) & & & \\ \hline
Cross	   &Resonance
         &$\sigma^R_h=676.277$    & $\tilde{\sigma}^R_h=0.266$
             &$\sigma^R_e=4.053$      & $\tilde{\sigma}^R_e=0.00158$ \\
section    &process
         &   &                    &   &                      \\ \cline{2-6}
($nb$)&Continuum
         &$\sigma^C_h=15.495$     &$\tilde{\sigma}^C_h=15.673$
	     &$\sigma^C_e=80.423$     &$\tilde{\sigma}^C_e=81.349$   \\
         &process
         &                    &   &                      \\ \hline \hline
\multicolumn{2}{c||}{Trigger efficiency $\sharp$} 
         &\multicolumn{4}{c}{Correction factor} \\ \hline
\multicolumn{2}{c||}{$\epsilon^{trg}_h=0.99924$}
         &\multicolumn{2}{c}{$\gamma_R = 0.999$} 
            &\multicolumn{2}{c}{$\gamma_C = 0.855$} \\ \hline \hline
\end{tabular}\\
{\footnotesize
$\sharp$: the trigger efficiency is obtained from Ref.~\cite{fucdtrg} written 
by Dr. Fu ChengDong. }
\end{center} 
\end{table}

$N^{TOT}_{\psip}$ is worked out to be either 14.05$\times 10^{6}$ for Fraction 
method or 14.04$\times 10^{6}$ for Normalization method.

\section{Error Analysis}
\subsection{Classification}
Formally, by the virtue of Eqs.~(\ref{totnmb1}) and (\ref{totnmb2}), 
the formula to calculate the total number can be written as
\begin{equation}
N^{TOT}_{\psip}=\frac{n^T_h}{{\cal B}_h \cdot \epsilon^R_h}
\cdot G_i~~~, 
\label{nterr}
\end{equation}    
where $G_i$ is a correction factor defined as
\[
G_i=\left\{
\begin{array}{ll}
G_1={\displaystyle \frac{1}{1+f}}~~, & \mbox{for the fraction method;} \\
G_2=(1- F_T \cdot M_{eh})~~,         & \mbox{for the normalization method.} 
\end{array}
\right.
\]
The error of $N^{TOT}_{\psip}$ comes from the components of
Eq. (\ref{nterr}), such as $n^T_h$, $\epsilon^R_h$, ${\cal B}_h$, and 
$G_i$, the error of which will be discussed one by one.

\subsection{Uncertainty of selected number $m^T_h$}
For the selected number\footnote{Hereafter the selected number $m$ instead of 
pure number $n$ is used in the error analysis, and the uncertainty for such 
substitution is rather small and is to be discussed afterwards.} $m^T_h$, there are three sources of uncertainty\footnote{Hereafter the symbol $\nu$ denotes 
relative error.}:
\begin{enumerate}
  \item {\bf Fitting uncertainty}\\
The uncertainty of fitted number could be obtained from the corresponding
error of fitting parameters which are used to calculate the number (refer
to Table~\ref{nmbeff}), that is
$$
\nu_{fit} (m^T_h) = \frac{\Delta m^{fit}_h}{m^{fit}_h}
                  = 0.11 \% $$ 
  \item {\bf Statistic uncertainty}\\
According to statistic principle, 
$$ \nu_{sta} (m^T_h) = \frac{1}{\sqrt{m^T_h}}
                     = 0.03 \% $$ 
  \item {\bf Selection uncertainty}\\
According to the study of section 1, the selection
uncertainty reflects the inconsistency between data ($m^T_h$) and 
Monte Carlo ($\epsilon^R_h$), so the uncertainty of $\epsilon^R_h$ is also
included in this term which is
$$ \nu_{sel} \left( \frac{m^T_h}{\epsilon^R_h} \right) 
         = 3.77 \% ~~.$$
  \item {\bf Effect due to beam energy fluctuation}\\
As it is mentioned in footnote \ref{ftnt_par}, the data are actually taken
within an enegy range (refer to the sketch description in Fig.~\ref{hadraw}), so
the ratio between $\sigma^C_h/\sigma^R_h$ will vary with the actual beam energy 
which may be different for different beam-injection. Usually, each 
beam-injection includes 3$-$5 runs. As an estimation, all
runs are grouped with every 3, 4 or 5 runs, then the ratios of selected hadron
number ($n^{T}_h$) to that of $\EE$ number ($n^{T}_e$) are worked out,
which denoted as $r_{he}(i)$ with $i$ indicating the grouped run number.
Fig.~\ref{sigsh} shows the $r_{he}$ distribution for different grouped-run 
number. The maximum value of $r_{he}(i)$ corresponds to the peak cross 
section\footnote{Notice 
$$ \sigma_h = \frac{n_h}{L \cdot \epsilon_h}~~, \mbox{~~~~and~~~~~~} 
          L = \frac{n_e}{ \sigma_e \cdot \epsilon_e}~~, $$
then
$$ \sigma_h = \frac{\sigma_e \cdot \epsilon_e}{\epsilon_h} \cdot 
              \frac{n_h}{n_e}~~, \mbox{~~~~or~~~~~}
  \sigma_h \propto \frac{n_h}{n_e}~~. $$	}. 
Taking the experimental statistic fluctuation into account, the value 10 is 
adopted as the position of the peak cross section. The uncertainty from the 
beam energy fluctuation effect is estimated as follows
$$\nu_{exp} = \frac{n^T_h/(1+f)}{\sum\limits_i n^T_h(i)/[1+f(i)]}~~,$$
with
$$f(i) =  \frac{\epsilon^C_h \cdot \sigma^C_h}{\epsilon^R_h \cdot \sigma^R_h}
  \cdot \frac{10}{r_{he}(i)}~~.$$
For different grouped-run cases, $\nu_{exp}$ is almost the same,
which is 0.23 \%.

\begin{figure}[htb]
\begin{minipage}{5.0cm}
\centerline{\psfig{file=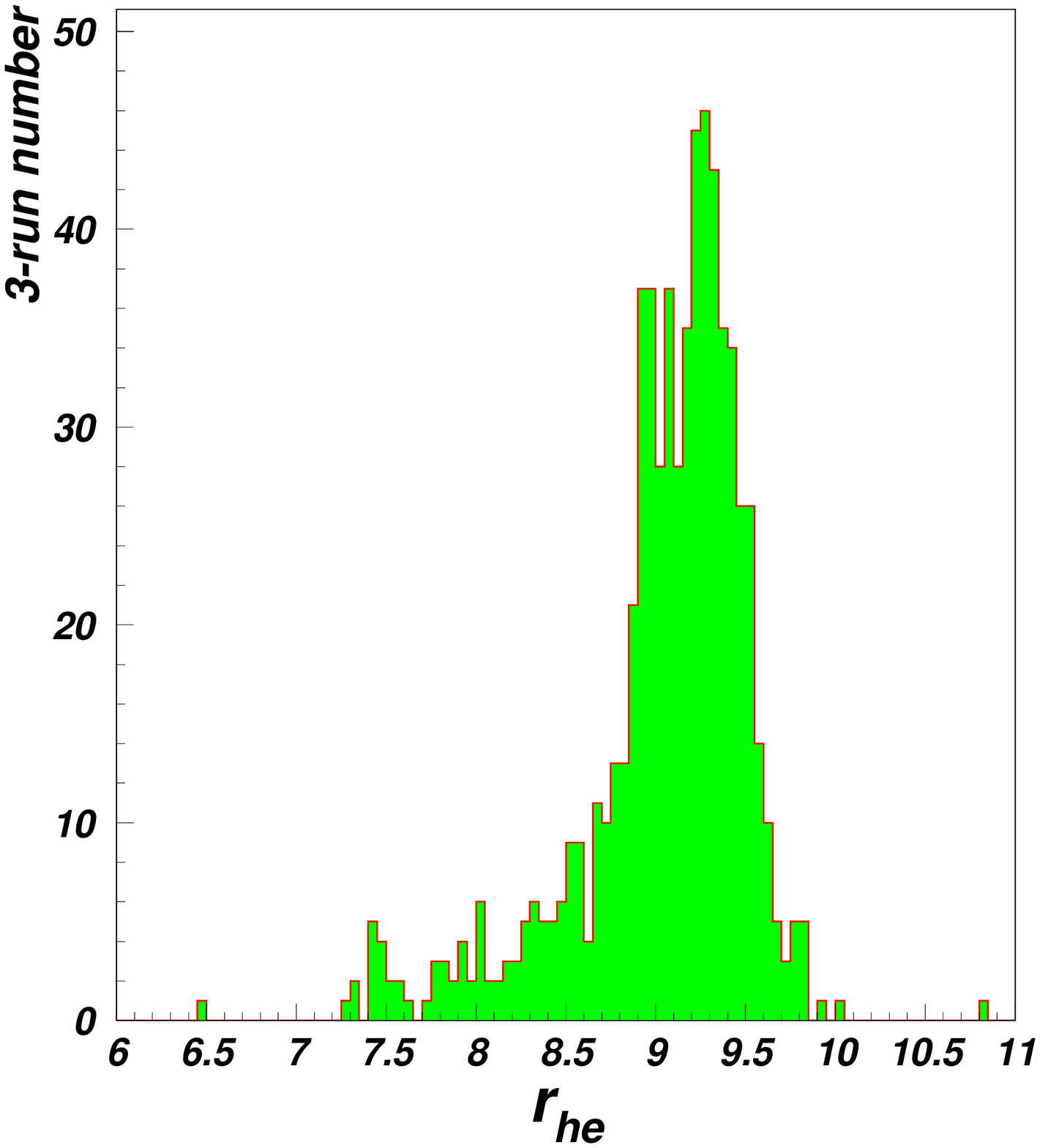,height=5.5 cm ,width=5 cm}}
\center (a) grouped with 3 runs 
\end{minipage}
\hskip 0.4cm
\begin{minipage}{5.0cm}
\centerline{\psfig{file=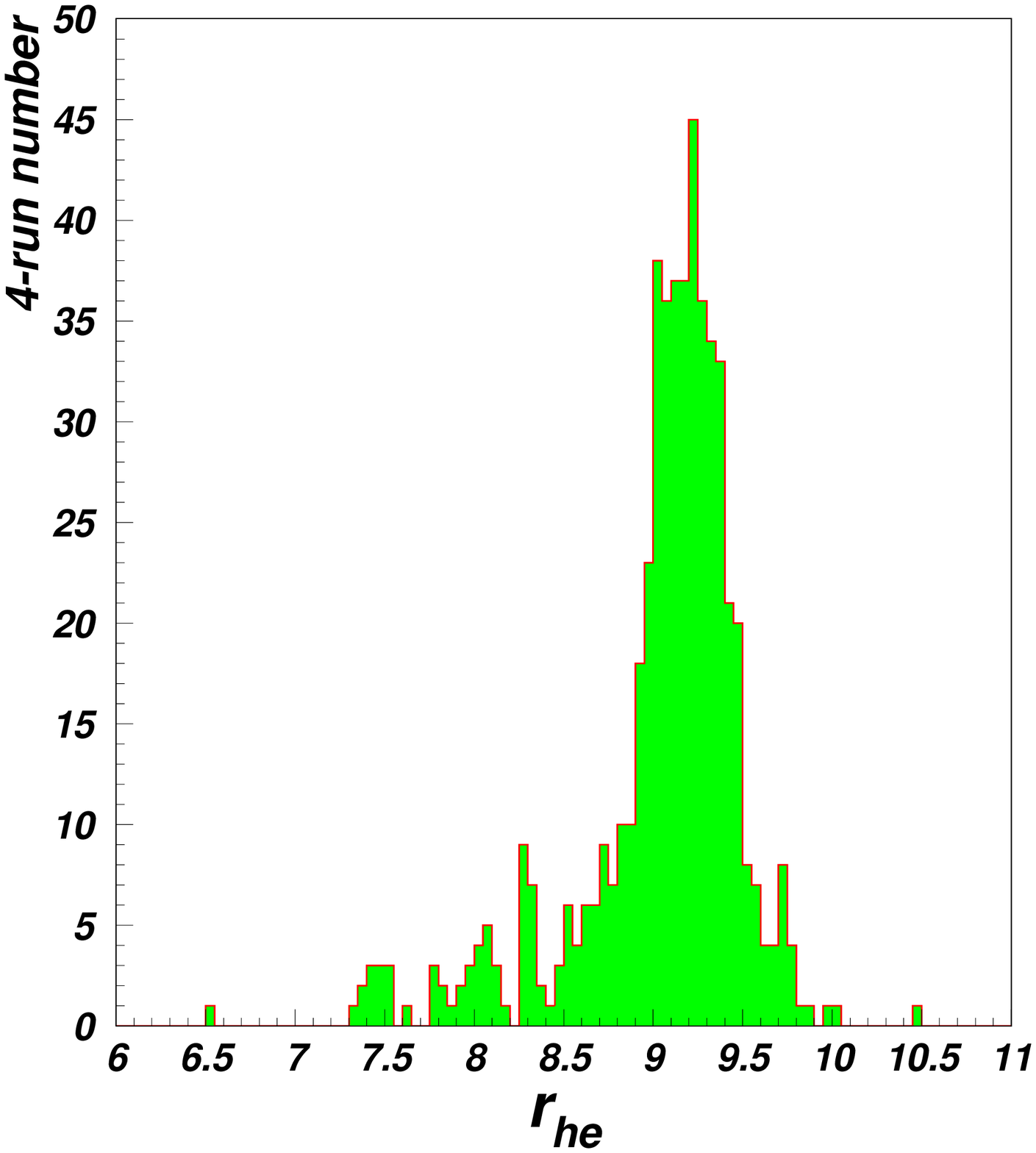,height=5.5 cm ,width=5 cm}}
\center (b) grouped with 4 runs
\end{minipage}
\hskip 0.4cm
\begin{minipage}{5.0cm}
\centerline{\psfig{file=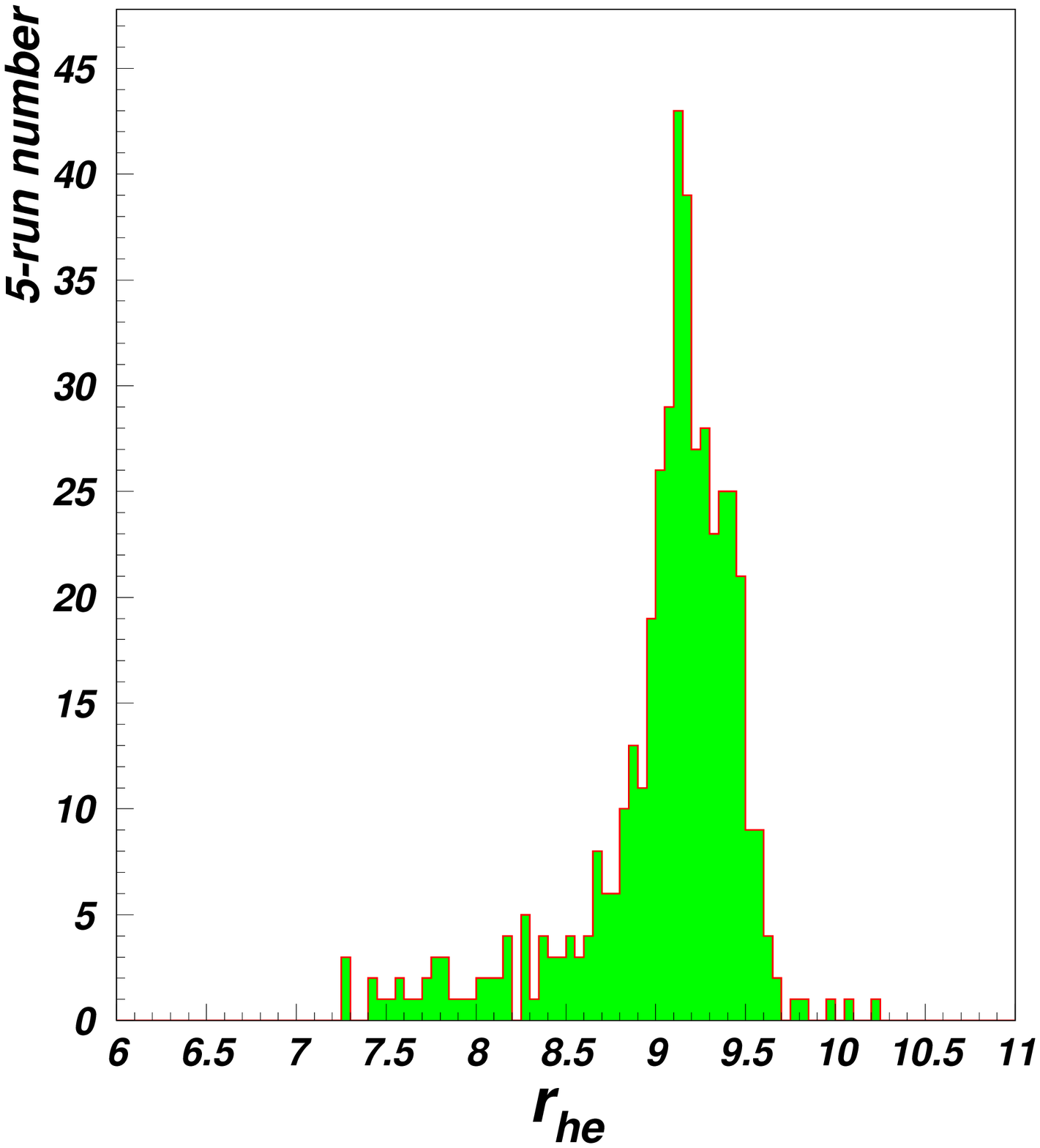,height=5.5 cm ,width=5 cm}}
\center (c) grouped with 5 runs
\end{minipage}
\caption{\label{sigsh}$r_{he}$ distributions for different grouped-run number.}
\end{figure}
\end{enumerate}  

\subsection{Uncertainty of branching ratio 
${\cal B}_h$}
There are two ${\cal B}_h$ values, one from PDG2002 and the other from
$\psip$ scan experiment:
\[
\begin{array}{rcl}
  {\cal B}_h(\mbox{PDG}2002)  & = &
(98.10 \pm 0.30) ) \% ~~, \\
 {\cal B}_h(\psip \mbox{ scan})& = &
(97.79 \pm 0.15 )\% ~~.
\end{array}
\]

In the total number calculation the ${\cal B}_h$ value form PDG2002 is used. 
However as a conservative estimation, the difference between above two
values is used as the uncertainty of ${\cal B}_h$, that is
$$
 \nu ( {\cal B}_h ) = 0.32\%~~.
$$

\subsection{Uncertainty of correction factor $G_i$}

For $G_1$, the uncertainty due to different $f$ could be calculated as follows
$$
 \nu (G_1)=
{\displaystyle  \left| \frac{N^{TOT}_{\psip}(f)-N^{TOT}_{\psip}(f^{\prime})}
{N^{TOT}_{\psip}(f)} \right| }= \left| 
\frac{\displaystyle \frac{1}{1+f}-\frac{1}{ 1+f^{\prime} } }
     {\displaystyle \frac{1}{1+f} } \right|
     =\left| \frac{\displaystyle f^{\prime}- f}
                  {\displaystyle 1 + f^{\prime}} \right| ~~.
$$
According to the definition of $f$, Eq.~(\ref{facf1}), the uncetainty of $f$ 
mainly comes from the statistics of $\epsilon^C_h$ and $\epsilon^R_h$, 
both of which are $1/\sqrt{50000}$. Notice $f=0.02179$ is a small quantity and 
the maximum difference between $f$ and $f^{\prime}$ is also small, so
$$ \nu (G_1) \approx |1-f| \cdot \Delta \approx 0.88 \%~~, $$
where $\Delta \equiv |f-f^{\prime}|_{max} = 2/\sqrt{50000}$.

For $G_2$, the uncertainty due to different $G_2$ could be calculated as follows
$$ \nu (G_2) ={\displaystyle
\left| \frac{N^{TOT}_{\psip}(G_2)- N^{TOT}_{\psip}(G_2^{\prime})}
            {N^{TOT}_{\psip}(G_2)} \right| }
=\left|  \frac{F_T \cdot M_{eh}- F^{\prime}_T \cdot M^{\prime}_{eh} }
      {1-F_T \cdot M_{eh}} \right|~~. $$
Notice $F_T(=0.932)$ approximates to one, so the difference of total number 
calculated with $F_T=1$ and with $F_T \neq 1$, is treated as the error from 
factor $F_T$, that is
$$ \nu (F_T)= {\displaystyle  \left| 
  \frac{M_{eh} \cdot (F_T-1) }{1-F_T \cdot M_{eh}}  \right| }
            = 0.20\%~~. $$
Notice $M_{eh}$ consists of two pairs of ratio, so the systematic
error of numerator and denominator will cancel automatically, only
the statistic error is left, which is
$$ \Delta M_{eh} = \sqrt{ \frac{1}{n^T_e}+\frac{1}{\tilde{n}^T_e}+
\frac{1}{\tilde{n}^T_h} +\frac{1}{n^T_h} }
=0.95\%~~, $$
so the error due to different $M_{eh}$ could be calculated as
$$ \nu (M_{eh})= {\displaystyle  \left| 
  \frac{F_T \cdot (M_{eh} - M^{\prime}_{eh}) }{1-F_T \cdot M_{eh}}  \right| } 
   \leq  {\displaystyle  \left|
     \frac{F_T \cdot \Delta M_{eh} }{1-F_T \cdot M_{eh}}  \right| }
   \approx 0.91\%~~. $$
Then the uncertainty for $G_2$ is
$$ \nu (G_2) = \sqrt{ \nu^2 (F_T) +\nu^2 (M_{eh})}
             = 0.94\%~~. $$

\subsection{Other uncertainty}
The other effects which could lead to the uncertainty include the correction 
factor $\gamma_C$, $\gamma_R$, trigger efficiency and so forth. All these 
uncertainties are regarded as less than 0.5\%.  In addition, Monte Carlo sample is used as input to test the bias of our method. The error from such bias is 
about 0.6\%.

\subsection{Summary}
Put all things together, the synthetic uncertainties are summarized in
Table~\ref{sumerr}. 
\begin{table}[htb]
\caption{\label{sumerr}Error summary.}
\doublerulesep 0.5pt
\center
\begin{tabular}{c|c||c} \hline \hline
\multicolumn{2}{c||}{ Source }       &  Uncertainty    \\ \hline
       $n^T_h$       & Fitting       &   0.11 \%      \\
                     & Statistic     &   0.03 \%      \\
($\& \epsilon^R_h$)  & Selection     &   3.77 \%       \\
                     &$E_{beam}$ fluctuation
		                     &   0.23 \%      \\ \hline

\multicolumn{2}{c||}{${\cal B}_h$}&   0.32 \%        \\ \hline
       $G_i$         & $G_1$      &   0.88 \%       \\
                     & $G_2$      &   0.94 \%      \\ \hline
\multicolumn{2}{c||}{Other}       &(0.5$\oplus$0.6) \% \\ \hline
       Total         & Method 1   &   3.97 \%       \\
                     & Method 2   &   3.98 \%    \\ \hline \hline
\end{tabular}\\
{\small Note: Method 1 for Fraction method; Method 2 for Normalizaion method.}
\end{table}

\section{Result and Discussion}

\setlength{\fboxrule}{0.5mm}
\setlength{\fboxsep}{4mm}

The final total number of $\psip$ event with corresponding error is

\setlength{\fboxrule}{0.8mm}
\setlength{\fboxsep}{5mm}
$$
N^{TOT}_{\psip}=  \left\{ 
\begin{array}{ll}
14.05 \times (1 \pm 3.97 \% )\times 10^6~~,\mbox{~~(Fraction Method)~;} \\
14.04 \times (1 \pm 3.98 \% )\times 10^6~~,\mbox{~~(Normalization Method)~.} 
\end{array} \right. 
$$

Notice Eqs.~(\ref{totnmb1}) and~(\ref{totnmb2}), two methods, the fraction 
method and the normalization method, correlated closely with each other and the difference between two methods for the central value is actually less than one 
per mille. Furthermore, the difference of the uncertainty for two methods is 
also fairly small. Therefore, the central value of the offline total number of 
$\psip$ event could be regarded as $14.0 \times 10^6$, and the uncertainty 
could conservatively be estimated as 4\%.

\vspace{1 cm}

{\em Thanks are due to Prof. J.~C.~Chen and 
Prof. D.~H.~Zhang for their help about Monte Carlo simulation.}



\begin{thebibliography}{**}
\thispagestyle{empty}
\bibitem{detector}
BAI Jing-Zhi et al. Nucl. Instr. Meth., 1999,{\bf A344}:319-334;\\
BAI Jing-Zhi et al. Nucl. Instr. Meth., 2001,{\bf A458}:627-637
\bibitem{psipscan}
BAI Jing-Zhi et al. Phys. Lett., 2002, {\bf B550}: 24-32
\bibitem{hadst}Mo Xiao-Hu, Study of Inclusive Hadronic Event, (2003.6),
BES Memo
\bibitem{jpsiscan}
BAI Jing-Zhi et al. Phys. Lett., 1995, {\bf B355}: 374-380.
\bibitem{PDG}Particle Data Group, Hagiwara K et al. ,
               Phys. Rev. 2002, {\bf D66}:010001
\bibitem{fucdtrg}FU Cheng-Dong, Measurement of the Trigger
Efficiency of $\psi^{\prime}$, (Feb. 27, 2003), BES Memo
\end{thebibliography}
\end{document}